\documentclass[conference,a4paper]{IEEEtran}
\IEEEoverridecommandlockouts
\usepackage{cite}
\usepackage{amsmath,amssymb,amsfonts}
\usepackage{algorithmic}
\usepackage{graphicx}
\usepackage{textcomp}
\usepackage{xcolor}

\usepackage{comment}

\usepackage{float}
\usepackage{subfig}
\usepackage{wrapfig}
\usepackage{framed,xcolor}
\usepackage{newfloat}
\usepackage[xcolor,framemethod=TikZ]{mdframed}
\usepackage{amsthm}
\usepackage{thmtools}
\usepackage{stmaryrd}
\usepackage{mathtools}
\usepackage{lineno}

\usepackage{graphicx}
\usepackage{amssymb,xfrac}
\usepackage{amsthm,amsmath,dsfont,bm}
\usepackage{thmtools}
\allowdisplaybreaks

\usepackage[algoruled,titlenotnumbered]{algorithm2e}
\usepackage{algorithmic}
\usepackage{wrapfig}
\usepackage{comment}
\usepackage{multirow}
\usepackage{amsmath,bm}
\usepackage{graphicx}
\usepackage{amsmath}

\newcommand{\dd}{\mathrm{d}}

\theoremstyle{theoremdd}

\newtheorem*{pf}{Proof}
\newtheorem{theorem}{Theorem}

\usepackage[a4paper,bindingoffset=0.2in,%
left=0.514in,right=0.5in,top=1.53in,bottom=0.75in,%
footskip=.25in]{geometry}

\begin{document}
	
\title{A User-Based Charge and Subsidy Scheme for Single O-D Network Mobility Management}

\author{Li~Li,~
	Dianchao~Lin$^{\dagger}$,~
	Saif~Eddin~Jabari,~
	\thanks{This work was supported by the NYUAD Center for Interacting Urban Networks (CITIES), funded by Tamkeen under the NYUAD Research Institute Award CG001 and by the Swiss Re Institute under the Quantum Cities™ initiative.}
	\thanks{Li Li is with the Department of Civil and Urban Engineering, New York University Tandon School of Engineering, NY, USA (email: ll3252@nyu.edu)}
	\thanks{Saif Eddin Jabari is with the Division of Engineering, New York University Abu Dhabi, UAE and the Department of Civil and Urban Engineering, New York University Tandon School of Engineering, NY, USA (email: sej7@nyu.edu)}
	\thanks{$^{\dagger}$ Corresponding author. Dianchao Lin is with the Department of Civil and Urban Engineering, New York University Tandon School of Engineering, NY, USA (email: dl3404@nyu.edu)}
}

\maketitle

\begin{abstract}
We propose a path guidance system with a user-based charge and subsidy (UBCS) scheme for single O-D network mobility management. Users who are willing to join the scheme (subscribers) can submit travel requests along with their VOTs to the system before traveling. Those who are not willing to join (outsiders) only need to submit travel requests to the system. Our system will give all users path guidance from their origins to their destinations, and collect a \emph{path payment} from the UBCS subscribers. Subscribers will be charged or subsided in a way that renders the UBCS strategy-proof, revenue-neutral, and Pareto-improving. A numerical example shows that the UBCS scheme is equitable and progressive.

\end{abstract}


\section{Introduction}
\label{S:intro}

Congestion pricing has been theoretically demonstrated to be an effective way of relieving traffic congestion in many pieces of literature since the pioneering work of \cite{pigou1920economics}. Comprehensive reviews can be found in \cite{yang2005mathematical, lawphongpanich2006mathematical, tsekeris2009design, lindsey2010reforming}. The classical approach to modeling the problem is to seek a tolling strategy under which a user equilibrium (UE) traffic assignment is the same as a system optimized (SO) assignment. Although congestion tolls work effectively in ideal situations, it faces both technical and political difficulties in the real world. Technically, according to almost all related studies, designing an optimal tolling strategy usually requires the authority to know about accurate and detailed demand information (OD demand matrix, value of time (VOT) distributions) \cite{wada2013hybrid}. However, due to the asymmetric information that exists between road managers and the users, it is almost impossible for the road managers to obtain that private information in the tolling framework. Politically, being enforced to pay for what was originally free makes the travelers regard the toll as a lump-sum tax and hence are resistant to the pricing policies. Equity impact is another important political issue. Congestion pricing has been believed to harm the poor people because they need to pay more due to inflexible schedules or are forced to switch to less desirable routes, transportation modes, or departure times \cite{taylor2010addressing}.

To deal with the technical issue about asymmetric information, some trial-and-error implementation methods have been proposed for a single road with unknown demand functions \cite{li2002role} (improved in \cite{wang2012bisection}), general network with unknown demand functions \cite{yang2004trial}, and general network with unknown demand \& cost functions \cite{yang2010road}. They tried to show that even with incomplete information, the proposed trial-and-error methods utilizing observable link flows can make the UE state converge to the SO state \cite{li2002role, wang2012bisection, yang2004trial}, or at least will lead to a general convergence \cite{yang2010road}. These methodologies, however, all assumed homogeneous travelers to simplify the modeling. Heterogeneous VOTs of individual travelers, which is a critical factor that influences their route choice and mode choice, are ignored. Hence the performance of these trial-and-error methods is still unknown in the real world where we surely have heterogeneous travelers.

To deal with the political issues, the \textit{toll and subsidy} (T\&S) schemes were designed to relieve public resistance and improve social equity. There include two types of T\&S schemes: the indirect two-stage schemes and the direct one-stage scheme. In the indirect two-stage schemes, the travelers are first charged nonnegative tolls for using the roads, and all the collected tolls are then redistributed to travelers in equal or unequal manners \cite{arnott1994welfare, eliasson2001road, guo2010pareto, mirabel2011bottleneck}. This makes the schemes revenue-neutral since it eliminates the financial transfer between the government and the travelers. In the direct one-stage scheme, the users are charged either positive tolls or negative tolls (subsidies) for using the roads or transportation modes \cite{bernstein1993congestion, adler2001direct, liu2009pareto, nie2010existence, xiao2014pareto, chen2012managing}. Usually, the sum of charges is equal to the sum of subsidies so that the schemes remain the revenue-neutral property. Apart from the objective of reaching SO under UE, the Pareto-improvement of the users' benefits has attracted significant attention when designing the T\&S scheme \cite{nie2010existence, guo2010pareto, xiao2014pareto}. A Pareto-improving scheme is a scheme that improves some travelers' benefits without sacrificing any others' benefits compared with the UE with no policy intervention. It is expected that if all travelers can benefit from the scheme (or at least some can benefit while others remain the same), the scheme will gain more public support. Efforts have shown that even if the Pareto-improvement is achieved, the T\&S schemes are usually regressive (benefit high-income travelers more than low-income travelers) \cite{wu2012design}. Hence they will widen the gap between the rich and poor.

In addition to the T\&S schemes, the \textit{tradable credit schemes} (TCS) have also been reputed to be an effective way to manage congestion in a revenue-neutral manner. In a TCS, credits are periodically (e.g., weekly, monthly) distributed by the authority to road users, and users are then required to pay a certain number of credits for using the links. Credits are universal within the network, while the credit charged may vary link by link. Credits can be traded among travelers so that people with extra credits can sell their credits to people with insufficient credits. The credit prices are determined through free trading. The TCS idea can be traced back to Dales's work in 1968 for water quality management \cite{dales1968pollution}. The concepts were then extended for vehicle emission control \cite{goddard1997using} and traffic management \cite{verhoef1997tradeable}.
A milestone was established in 2011 when \cite{yang2011managing} made rigorous quantitative modeling and analysis of the TCS for network mobility management. Following that, different efforts have been made to study the TCS from various aspects. A recent detailed review of TCS-related research can be found in \cite{xiao2019promoting}. 

The most straightforward advantage of TCS over direct congestion pricing is its revenue-neutral property, which helps reduce public skepticism. It has also been demonstrated that an O-D specific distribution of the initial credits can make the TCS Pareto-improving considering homogeneous travelers \cite{yang2011managing}. Researches considering travelers' different income levels have found that TCS can be more equitable and progressive (benefit low-income travelers more than high-income travelers) than congestion pricing \cite{wu2012design}. Despite those improvements in political aspects, the technical issue when it comes to implementation in practice is by no means well solved. The TCS still needs specific O-D demand and individual VOT information. Similar to congestion pricing, the trial-and-error methods have also been proposed to implement the TCS with unknown demand functions for both single road \cite{wang2012bisection} and general network \cite{wang2014trial}. VOT was assumed to be homogeneous and known in these studies. No effective implementation strategies for TCS have been developed for a more realistic case with unknown heterogeneous VOTs.

As an alternative to congestion pricing and TCS, there exist other kinds of economic instruments that seek to reduce traffic congestion and remove the asymmetric information issue. One example is the \textit{tradable network permits} (TNP) proposed in \cite{akamatsu2017tradable}.
In TNP, an operator issues time-dependent permits for each link, and only the permit holder has the right to use a pre-specified link at a pre-specified time. The operator can eliminate congestion by ensuring that the number of issued permits does not exceed the capacity of each link. Therefore, link travel times are assumed to be constant due to the absence of congestion. Hence users will be able to arrive at their purchased link on time. Permits can be distributed in a market selling scheme or a free distribution scheme. In a market selling scheme, travelers need to buy permits from the operator, which deviates the TNP from revenue-neutral. In a free distribution scheme, permits are freely distributed among travelers. Travelers can trade the permits in a permit market when their holding permits do not match their travel demands. According to the demonstration in \cite{akamatsu2007system, akamatsu2017tradable}, the TNP theoretically only requires the operator to know the capacity information of each link for determining the number of issued permits, and there is no need to know any demand information. 

In summary, pure congestion pricing is not revenue-neutral or equitable, although it can be Pareto-improving in certain special networks \cite{lawphongpanich2007pareto, lawphongpanich2010solving, song2009nonnegative}. T\&S scheme is revenue-neutral, can be Pareto-improving in some cases, but it is hard to be equitable. TCS is revenue-neutral, can be Pareto-improving in some cases, and it is more equitable than pricing and T\&S scheme. All of the three aforementioned schemes require detailed demand information (even with existing trial-and-error methods, the detailed VOT information is still needed), which makes their performance unknown in the real world with heterogeneous travelers. Even if we assume that the operators have access to ask travelers about their VOTs, no existing strategies ensure that the travelers will trustfully report their VOTs. In other words, the strategy-proofness of all these schemes is unknown. TNP is usually not revenue-neutral or equitable; its Pareto-improving property is unknown. However, it is more practical in the sense that no demand information is required. Moreover, it is proved to be strategy-proof under a properly designed agent-based auction mechanism \cite{wada2013hybrid}.

In light of the research gap and the inspirations from the literature, we propose a path guidance system with a user-based charge and subsidy (UBCS) scheme for single O-D network mobility management. Users who are willing to join the scheme (subscribers) can submit travel requests along with their VOTs to the system before traveling. Those who are not willing to join (outsiders) only need to submit travel requests to the system. Our system will give all users path guidance from their origins to their destinations, and collect a path payment (can be either positive or negative) from the UBCS subscribers. Subscribers will be charged or subsided in a way that renders the UBCS scheme strategy-proof, revenue-neutral, and Pareto-improving. Our numerical example also shows that the UBCS scheme is equitable and progressive, which means it benefits the poor more than the rich.

\section{Methodology}
\label{sec:method}

We assume a single O-D with fixed demand in this paper. Our approach consists of four steps, which we describe next.

\medskip

\textbf{Step 1: Link flow determination:} Our optimal link flow vector, $q^*=\{q^{*}_a\}$, is one that solves a classical single O-D SO problem, which (for the sake of completeness) is given as
\allowdisplaybreaks
\begin{align}
\mathrm{Minimize} & \quad \sum_{a \in A} q_a t_a(q_a) \label{p1_obj} \\
\mathrm{s.t.} & \quad \sum_{r \in R}\delta_{a,r} f_r  = q_a, ~~ a \in A \label{p1_c1} \\
& \quad \sum_{r \in R} f_r = d \label{p1_c2} \\
& \quad f_r \ge 0, ~~ r \in R. \label{p1_c3} 
\end{align}
Here $\delta_{a,r} = 1$ if path $r$ uses link $a$, and 0 otherwise, $d$ is fixed travel demand, $t_a(\cdot)$ is a travel time function associated with link $a$ (assumed to be a non-negative, convex, differentiable and monotonically increasing function with link volume) and $f_r$ is the flow on path $r$. Since the objective function and constraints for the above SO problem are convex, the optimal link flow vector $q^{*}$ is unique, while the optimal path flow vector $f^{*} = \{f^{*}_r\}$ is not necessarily unique.

\medskip

\textbf{Step 2: Path flow determination:} For the UBCS subscribers, we consider continuously distributed VOTs with bounded support and denote the individual VOT by $\beta$. We denote by $\tilde{d}$ the portion of $d$ that corresponds to subscribers.
We then divide 
the support of the VOT distribution into $M$ intervals of equal length $\Delta \beta$.
Let $\tilde{d}^m$ denote the number of subscribers whose VOTs are within the $m$th interval, $\beta^m$ the average VOT of all $\tilde{d}^m$ users within the $m$th interval, and $\tilde{f}^{m}_r$ the flow of subscribers on path $r$ with VOTs within the $m$th interval. 
The cost-minimizing path flows corresponding to subscribers can be determined by producing a distribution of subscriber demands to routes based on their VOT (weighted by VOT), but preserving SO flows. Constraint \eqref{p2_c1} below assigns the link flow to subscribers in proportion to their volume so that resources are fairly distributed among subscribers and outsiders.
\begin{align}
\mathrm{Minimize} & \quad \sum_{m=1}^{M}\sum_{r \in R} \beta^m \tilde{f}^m_r T_r \label{p2_obj} \\
\mathrm{s.t.} & \quad \sum_{m=1}^{M}\sum_{r \in R} \delta_{a,r} \tilde{f}^m_r  = q^{*}_a \frac{\tilde{d}}{d}, ~~ a \in A \label{p2_c1} \\
& \quad \sum_{r \in R} \tilde{f}^m_r = \tilde{d}^m, ~~ m = 1,\hdots,M  \label{p2_c2} \\
& \quad \tilde{f}^{m}_r \ge 0, ~~ r \in R. \label{p2_c3} 
\end{align}
Here $T_r$ is the travel time of path $r$ calculated as
\begin{equation}
T_r = \sum_{a \in A} \delta_{a,r} t_a(q^{*}_a).
\end{equation}
Let $\hat{f_r}$ denote the flow of outsiders on path $r$, and $\hat{d} = d - \tilde{d}$ the outsiders' demand. Based on $q^{*}$ (in the first step) and the payment path flow vector $\tilde{f}^{m*} = \{\tilde{f}^{m*}_r\}_{r \in R}$ (the second step), it can be readily verified that the path flow of outsiders is given by
\begin{equation}
\hat{f}^*_r = \sum_{m=1}^{M} \frac{\hat{d}}{\tilde{d}} \tilde{f}^{m*}_r.
\end{equation}

\medskip

\textbf{Step 3: Individual user assignment:} 
%
%
Let $(r_{(1)}, \cdots, r_{(|R|)})$ be a rearrangement of path set $R$ so that $T_{r_{(1)}} \ge \cdots \ge T_{r_{(i-1)}} \ge T_{r_{(i)}} \ge \cdots \ge T_{r_{(|R|)}}$, and with slight abuse of notation we define a corresponding partition of the support of the VOT distribution $\inf \beta = \beta_0 \le \beta_1 \le \hdots \le  \beta_{|R|} = \sup \beta$, where\footnote{The partition points $\{\beta_i\}$ differ from the discrete VOTs $\{\beta^m\}$ in step 2.}
\begin{equation}
\int_{\beta_0}^{\beta_i} p(b) \dd b =\frac{1}{\tilde{d}}\sum_{j=1}^{i} \tilde{f}^*_{r_{(j)}},
\label{E:VOTi}
\end{equation}
where $p$ is probability density function of VOT $\beta$.  Each interval $(\beta_{i-1}, \beta_i]$ gives the lower and upper bound VOT associated with path $r_{(i)}$ with travel time $T_{r_{(i)}}$. Hence subscribers with VOT $\beta \in (\beta_{i-1}, \beta_i]$ will be assigned to path $r_{(i)}$.  We denote by $\rho_i$ the likelihood that an outsider is assigned to route $r_{(i)}$, which is given by
\begin{equation}
\label{rho}
\rho_i = \frac{\tilde{f}^{*}_{r_{(i)}}}{\tilde{d}} = \frac{\hat{f}^{*}_{r_{(i)}}}{\hat{d}}.
\end{equation}


%

\medskip

\textbf{Step 4: Subscribers' payment determination:} 
We use the following equation to determine the UBCS subscribers' payment on each path $r_{(i)}$, $i=1,\hdots,|R|$:
\begin{multline}
\label{pay}
P_i = \sum_{h=1}^{i-1} \rho_h \sum_{g=h+1}^{i}(T_{r_{{(g-1)}}}-T_{r_{{(g)}}})\beta_{g-1} - \\
\sum_{h=i+1}^{|R|} \rho_h \sum_{g=i+1}^{h}(T_{r_{{(g-1)}}}-T_{r_{{(g)}}})\beta_{g-1}.
\end{multline}

According to \eqref{pay}, subscribers on the same path $r_{(i)}$ make the same payment, even if they have different VOTs.  

\begin{theorem}
	\label{Theorem: sp}
	The payment determined by \eqref{pay} is the unique strategy that is both strategy-proof and revenue-neutral.
\end{theorem}

\begin{pf}
	Assume that $\exists \beta \in (\beta_{i-1},\beta_i]$ in path $i \in R_+$, and $\exists \beta' \in (\beta_{j-1},\beta_j]$ in path $j \in R_+$.  Without loss of generality, we just make $\beta < \beta'$ here.
	
	\textbf{1) In the case of $i = j$}
	
	If $\beta$ is the true VOT and $\beta'$ is the declared one, a strategy-proof mechanism should require additional cost brought by lying to be non-negative:
	
	\begin{multline}
		C(\beta' |\beta)  - C(\beta |\beta) = T_{r_{(j)}} \beta + P_{\beta',j} - T_{r_{(i)}} \beta - P_{\beta,i} \\ \overset{i=j}=  P_{\beta',i} - P_{\beta,i} \ge 0. \label{E:C_dif1}
	\end{multline}
	Where $P_{\beta,i}$ is payment of the user with declared VOT $\beta$ (in path \textit{i}).
	$C$ represents the cost.
	If $\beta'$ is the true VOT and $\beta$ is the declared one, we have
	
	\begin{multline}
		C(\beta |\beta')  - C(\beta' |\beta') = T_{r_{(i)}} \beta' + P_{\beta,i} - T_{r_{(j)}} \beta' - P_{\beta',j} \\ \overset{i=j}=  P_{\beta,i} - P_{\beta',i} \ge 0. \label{E:C_dif2}
	\end{multline}
	
	Based on inequalities \eqref{E:C_dif1} and \eqref{E:C_dif2}, we have $P_{\beta,i} = P_{\beta',i}$. It means that, \textbf{the payment for subscribers with different VOTs on the same path should be the same}. For simplification, we use  $P_i $ to represent the payment on path $i$ in the following content.
	
	\textbf{2) In the case of $i \not= j$}
	
	Since $\beta < \beta'$, we have $i < j$, hence $T_{r_{(i)}} \ge T_{r_{(j)}}$.	
	Similar to constraints \eqref{E:C_dif1} and \eqref{E:C_dif2},we have
	
	\begin{equation}
		P_j - P_i \ge (T_{r_{(i)}} - T_{r_{(j)}}) \beta_i. \label{E:P_dif1}
	\end{equation}
	
	\begin{equation}
		P_j - P_i \le (T_{r_{(i)}} - T_{r_{(j)}}) \beta_{j-1}. \label{E:P_dif2}
	\end{equation}
	
	Based on \eqref{E:P_dif1} and \eqref{E:P_dif2}, we can have a unique solution:
	\begin{equation} 
		P_j - P_i = \sum_{k=1}^{j-i}(T_{r_{({i+k-1})}} - T_{r_{({i+k})}}) \beta_{i+k-1}. \label{E:P_dif4}
	\end{equation}
	
	To make the mechanism revenue neutral, we need one more equation:
	\begin{equation} 
		\sum_{i=1}^{|R|}\rho_i P_i = 0, \label{E:rhoP}
	\end{equation}
	Solving the equations yields the unique solution $P_i$ ($i = 1, \cdots, |R_+|$):
	
	\begin{multline} 
		P_i = \sum_{h=1}^{i-1} \rho_h \sum_{g=h+1}^{i}(T_{r_{{(g-1)}}}-T_{r_{{(g)}}})\beta_{g-1} - \\
		\sum_{h=i+1}^{|R|} \rho_h \sum_{g=i+1}^{h}(T_{r_{{(g-1)}}}-T_{r_{{(g)}}})\beta_{g-1}. \label{E:Pi}
	\end{multline}
	This completes the proof. $\square$
	
\end{pf}

Theorem \ref{Theorem: sp} states that the UBSC scheme is unique, and that under the UBSC scheme no subscribers are incentivized to report untruthful VOTs (or keep their VOT information private). Note that VOT is collected before each trip, hence one person may have different VOTs in different trips, depending on the trip purpose and other circumstances. Travelers are responsible for evaluating their specific VOT in each trip, and the UBCS scheme will motivate them to report the true evaluated VOT. Meanwhile, the UBCS scheme is revenue neutral: the total payments made are equal to the subsidies made in the system.

It is worth noting that, we do not explicitly introduce a path-differentiated payment, it is rather natural outcome of our strategy-proof approach. This fundamentally differs from path-differentiated pricing in the literature which explicitly employ a unique toll for each path \cite{zangui2013differentiated, zangui2015path}. However, since our model produces a path-differentiated payment as output, the research on how to efficiently implement path tolls using automated vehicle identification sensors given a path toll scheme \cite{zangui2015sensor} can be also used when implementing our UBCS in the real world.

\begin{theorem}
	\label{Theorem:PI}
	Joining the UBCS scheme is Pareto-improving compared with either quitting the UBCS scheme or being in the UE without any policy intervention.
\end{theorem}

\begin{pf}
	Suppose there is a user with VOT $\beta \in (\beta_{i-1},\beta_i]$. If he/she is a UBCS joiner, we can calculate his/her cost $C(\beta)$ as	
	\begin{equation} 
	C(\beta) = T_{r_{{(i)}}} \beta + P_i. \label{E:C}
	\end{equation}
	If he/she is a UBCS quitter, we can calculate his/her cost $C_Q(\beta)$ as
	\begin{equation} 
    C_Q(\beta) = \sum_{h=1}^{|R|}\rho_h T_{r_{{(h)}}} \beta. \label{E:C_Q}
	\end{equation}
	If no policy is implemented, under the UE condition, his/her cost $C_{UE}$ can be calculated as
	\begin{equation} 
	C_{UE}(\beta) = T_{UE} \beta, \label{E:C_UE}
	\end{equation}
	where $T_{UE}$ is the universal travel time for all users under UE.
	
	Combining Equations \eqref{E:Pi} to \eqref{E:C_Q} and $\sum_{h=1}^{|R|}\rho_h = 1$, we have
	\begin{multline} 
	C_Q(\beta)  - C(\beta) =
	-\sum_{h=1}^{i-1} \rho_h \sum_{g=h+1}^{i}(T_{r_{{(g-1)}}}-T_{r_{{(g)}}})\beta_{g-1} +  \\
	 \sum_{h=1}^{|R|}\rho_h T_{r_{{(h)}}} \beta - T_{r_{{(i)}}} \beta +
	\sum_{h=i+1}^{|R|} \rho_h \sum_{g=i+1}^{h}(T_{r_{{(g-1)}}}-T_{r_{{(g)}}})\beta_{g-1}\\
	= \sum_{h=1}^{i-1}\rho_h (T_{r_{{(h)}}}-T_{r_{{(i)}}}) \beta -
	\sum_{h=1}^{i-1} \rho_h \sum_{g=h+1}^{i}(T_{r_{{(g-1)}}}-T_{r_{{(g)}}})\beta_{g-1} + \\
	\sum_{h=i+1}^{|R|}\rho_h (T_{r_{{(h)}}}-T_{r_{{(i)}}}) \beta + 
	\sum_{h=i+1}^{|R|} \rho_h \sum_{g=i+1}^{h}(T_{r_{{(g-1)}}}-T_{r_{{(g)}}})\beta_{g-1} \\
	= \sum_{h=1}^{i-1} \rho_h [(T_{r_{{(h)}}}-T_{r_{{(i)}}}) \beta - \sum_{g=h+1}^{i}(T_{r_{{(g-1)}}}-T_{r_{{(g)}}})\beta_{g-1}] + \\
	\sum_{h=i+1}^{|R|} \rho_h [(T_{r_{{(h)}}}-T_{r_{{(i)}}}) \beta + \sum_{g=i+1}^{h}(T_{r_{{(g-1)}}}-T_{r_{{(g)}}})\beta_{g-1}].
		\label{E:C_dif}
	\end{multline}
	
	For any $1 \leq h \leq i-1$, there are  $\beta_h \leq \cdots \leq \beta_{i-1} \leq \beta$, and $T_{r_{{(h)}}} \geq T_{r_{{(i)}}}$:
	\begin{multline} 
	(T_{r_{{(h)}}}-T_{r_{{(i)}}}) \beta - \sum_{g=h+1}^{i}(T_{r_{{(g-1)}}}-T_{r_{{(g)}}})\beta_{g-1} \geq \\
	(T_{r_{{(h)}}}-T_{r_{{(i)}}}) \beta - \beta_{i-1}\sum_{g=h+1}^{i}(T_{r_{{(g-1)}}}-T_{r_{{(g)}}}) = \\
	(T_{r_{{(h)}}}-T_{r_{{(i)}}})(\beta-\beta_{i-1}) \geq 0
	\end{multline} 
	
	For any $i+1 \leq h \leq|R|$, there are  $\beta \leq \beta_i \leq \cdots \leq \beta_h$, and $T_i \geq T_h$:
	\begin{multline} 
	(T_{r_{{(h)}}} - T_{r_{{(i)}}}) \beta + \sum_{g=i+1}^{h}(T_{r_{{(g-1)}}}-T_{r_{{(g)}}})\beta_{g-1} \geq \\
	(T_{r_{{(h)}}} - T_{r_{{(i)}}}) \beta + \beta_i \sum_{g=i+1}^{h}(T_{r_{{(g-1)}}}-T_{r_{{(g)}}}) = \\
	(T_{r_{{(i)}}} - T_{r_{{(h)}}}) (\beta_i-\beta)  \geq 0
	\end{multline} 
	
	Therefore, $C_Q(\beta)  - C(\beta) \geq 0$. Hence joining the UBCS is Pareto-improving compared with quitting it for all users with different VOTs.
	
	Combining \eqref{E:C_Q} and \eqref{E:C_UE}, we have
	\begin{equation}
	C_{UE}(\beta) - C_Q(\beta) = \beta(T_{UE} - \sum_{h=1}^{|R|}\rho_h T_{r_{{(h)}}}) 
	\end{equation}
	Since $T_h$ is based on the time-based SO link flow solution from the first step, we can find that the expected travel time of the quitters is no longer than the UE travel time, which means $T_{UE} \geq \sum_{h=1}^{|R_+|}\rho_h T_h$. Hence $C_{UE}(\beta) \geq C_Q(\beta) \geq C(\beta)$, meaning that  joining the UBCS scheme is Pareto-improving compared with UE without any policy intervention.
	
	This completes the proof. $\square$	
\end{pf}

According to Theorem \ref{Theorem:PI}, the guidance system is attractive to current users since joining the system (either being a subscriber or an outsider) will reduce their travel costs in expectation. While in the system, joining the UBCS scheme would be attractive to users since it will further reduce their expected costs. Such Pareto-improvement property makes the UBCS scheme active without forcing anyone to join it.

\section{Numerical example}
\label{sec:numerics}

\begin{figure}[h!]
		\centering
	\resizebox{0.355\textwidth}{!}{%
		\includegraphics{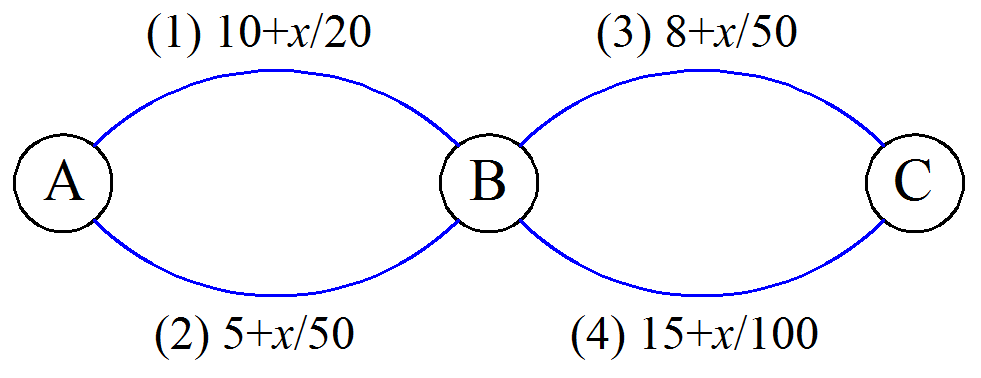}}
	\caption{\textbf{ Network structure}} 
	\label{F:network}
\end{figure}

\begin{figure}[h!]
	\centering
\resizebox{0.27\textwidth}{!}{%
	\includegraphics{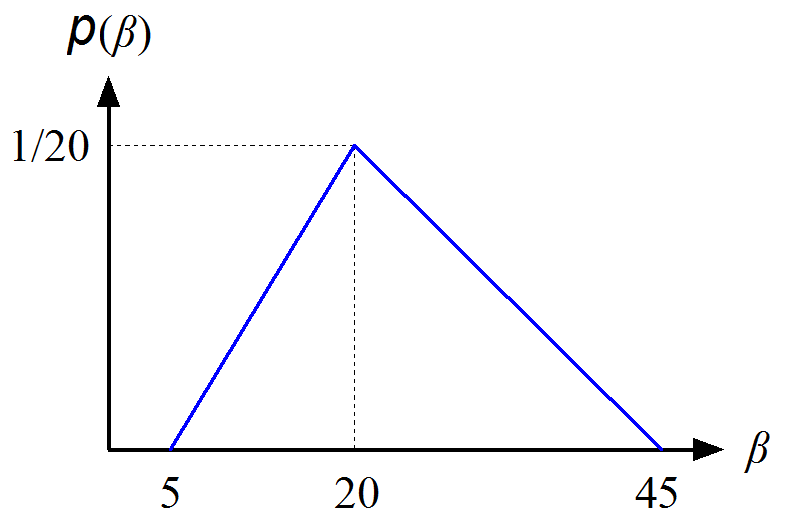}}
\caption{\textbf{VOT distribution}} 
\label{F:VOT}
\end{figure}

Consider the small network in Fig. \ref{F:network}; the number inside the bracket is the link number, and the function beside it is the link travel time (min) function of the link flow. Note that the linearly increasing functions are used here for simplicity, more complicated and practical functions can be used in real world application. The OD demand from A to C is 1000. We assume 800 users are UBCS subscribers, and the remaining 200 are outsiders. The PDF of VOT for the subscribers is assumed to be triangular as shown in Fig. \ref{F:VOT}.

\begin{table}[h!]
	\centering
	\caption{\textbf{Link flows under UE and SO}}
	\begin{tabular}{|c c c c c|}
		\hline
		\multicolumn{1}{|c|}{\textbf{Link}}
		& (1) & (2) & (3) & (4) \\
		\hline
		\multicolumn{1}{|c|}{\textbf{UE flow}}
		&214& 786 & 567 & 433 \\
		\multicolumn{1}{|c|}{\textbf{UE time} (min)}
		& 20.7 & 20.7 & 19.3 & 19.3 \\
		\multicolumn{1}{|c|}{\textbf{SO flow}}
		& 250 & 750 & 450 & 550 \\
		\multicolumn{1}{|c|}{\textbf{SO time} (min)}
		& 22.5 & 20 & 17 & 20.5 \\
		\hline
	\end{tabular}
	\label{T:UESOresults}
\end{table}

\begin{table}[h!]
	\centering
	\caption{\textbf{Results for UBCS subscribers and outsiders}}
	\begin{tabular}{|c|c|c|c|c|}
		\hline
		\multicolumn{1}{|c|}{\textbf{Path}}
		&(1) + (3) & (1) + (4) & (2) + (3) & (2) + (4) \\
		\hline
		\multicolumn{1}{|c|}{\textbf{Time} (min)}
		&39.5& 43 & 37 & 40.5 \\
		\multicolumn{1}{|c|}{\textbf{Subscribers}}
		&0& 200 & 360 & 240 \\
		\multicolumn{1}{|c|}{\textbf{Outsiders}}
		&0& 50 & 90 & 60 \\
		\multicolumn{1}{|c|}{\textbf{VOT} (\$/h)}
		&-& $(5.0,17.2]$ & $(31.6,45.0]$ & $(17.2,31.6]$ \\
		\multicolumn{1}{|c|}{\textbf{Payment} (\$)}
		&-& $-1.37$ & $1.19$ & $-0.65$ \\
		\hline
	\end{tabular}
	\label{T:UBCSresults}
\end{table}

\begin{figure}[h!]
	\centering
	\resizebox{0.3525\textwidth}{!}{%
		\includegraphics{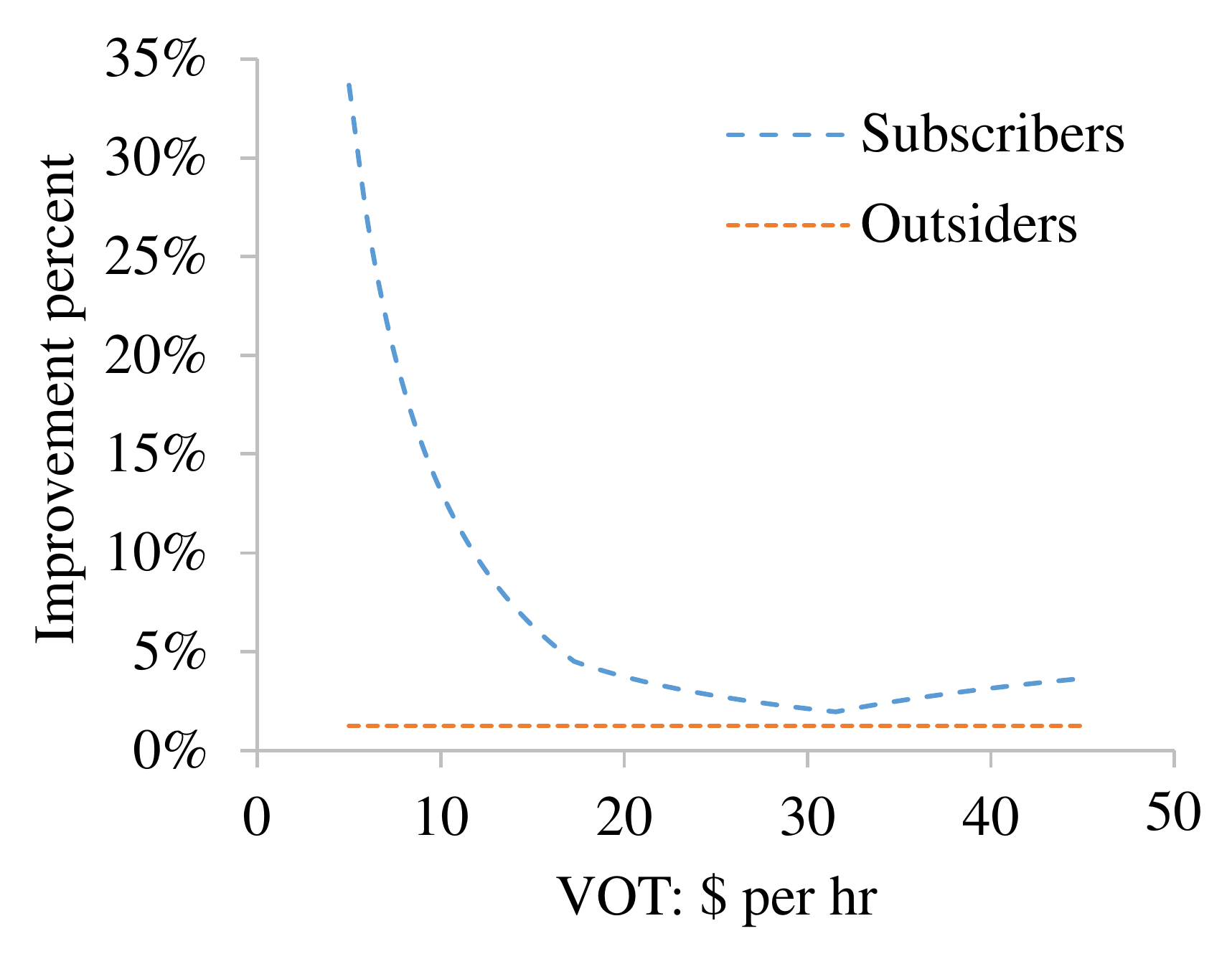}}
	\caption{\textbf{Improvement of UBCS compared with the UE with no policy intervention.}} 
	\label{F:additional_cost}
\end{figure}

The link flows from UE and SO are shown in TABLE \ref{T:UESOresults}. Average travel time in UE and SO is 40.1 min and 39.6 min, respectively. Through UBCS, we can guide users to use allocated paths, and make corresponding payments.
The optimal travel times for all paths, subscriber and outsider path flows, the VOT bounds, and payments for each path are shown in TABLE \ref{T:UBCSresults}. According to the UBCS scheme, 360 subscribers with highest VOTs are guided to the fastest path with travel time of 37 min. Each of them will be charged \$1.19. 240 Subscribers with middle VOTs are guided to the path with travel time of 40.5 min, and each of them will receive a subsidy of \$0.65. The rest subscribers with lowest VOTs are guided to the path with longest travel time of 43 min, and will receive a subsidy of \$1.37.

We also show the percent improvement to UBCS outsiders and subscribers compared with UE conditions without any policy intervention. Fig. \ref{F:additional_cost} shows the result for users with different VOTs. We find that subscribers with low VOT ($\leqslant$ \$18/h) benefit the most from joining the UBCS scheme, which is much higher than the subscribers with mid-to-high VOTs. Subscribers with the lowest VOTs can enjoy improvements as high as 34\%. This demonstrates that our UBCS scheme is equitable and progressive, meaning it benefits the poor more than the rich. Besides, outsiders can also benefit from the system compared with UE, while the improvement percentage remains the same for all outsiders with different VOTs.

\section{Conclusion}
\label{sec:conc}

Based on a single O-D network with fixed demand, we proposed a path guidance system with a user-based charge and subsidy (UBCS) scheme for network traffic management. In our system, users need to submit their travel request to road operators in advance. Moreover, if they are UBCS subscribers, they also need to report their VOT to operators. The system optimizes all requests and then returns path guidance to each user. A charge or subsidy will be made to the user based on their allocated path. We theoretically proved that the UBCS is strategy-proof (encourages the subscribers to report their VOTs truthfully), Pareto-improving (no one is worse off compared with the UE with no policy intervention), and revenue-neutral. Our numerical example shows that the UBCS is also equitable and progressive. It can help reduce the gap between poor and rich, implying it is more likely to gain public support. The essence of the UBCS scheme is the win-win utility transfer between different users so that all users can benefit from the scheme. Similar ideas have also been employed in the lane change scenario \cite{lin2019pay, lin2019pay1} and single intersection control scenario \cite{lin2019transferable, lin2020pay} in our earlier work. Future studies can include multi O-D analysis and elastic demand scenarios.

\bibliographystyle{IEEEtran}
\bibliography{refs.bib}

\end{document}